\documentclass[final, runningheads]{llncs}

\def\anonymous{no}

\usepackage{ifthen}

\usepackage{amsmath,amssymb}
\usepackage{enumitem}

\usepackage{color}
\usepackage[usenames,dvipsnames]{xcolor}
\usepackage{xspace}

\usepackage{graphicx}
\usepackage{subcaption}
\usepackage[unicode,colorlinks=true,urlcolor=blue,citecolor=red]{hyperref}%
\usepackage[capitalise, noabbrev]{cleveref}
\usepackage{csquotes}

\newcommand{\cro}[1]{\ensuremath{\operatorname{\overline{cr}}(#1)}\xspace}

\newcommand{\M}[1]{\ensuremath{\operatorname{M}_{#1}^{\vee}}\xspace}

\spnewtheorem{observation}[theorem]{Observation}{\bfseries}{\itshape}
\spnewtheorem{ques}{Question}{\bfseries}{\itshape}
\usepackage{thmtools, thm-restate}

\newenvironment{claimproof}[1]{\underline{Proof of Claim:}\space#1}{\leavevmode\unskip\penalty9999 \hbox{}\nobreak\hfill\quad\hbox{$\blacksquare$}}

\renewcommand{\orcidID}[1]{\href{https://orcid.org/#1}{\includegraphics[scale=.03]{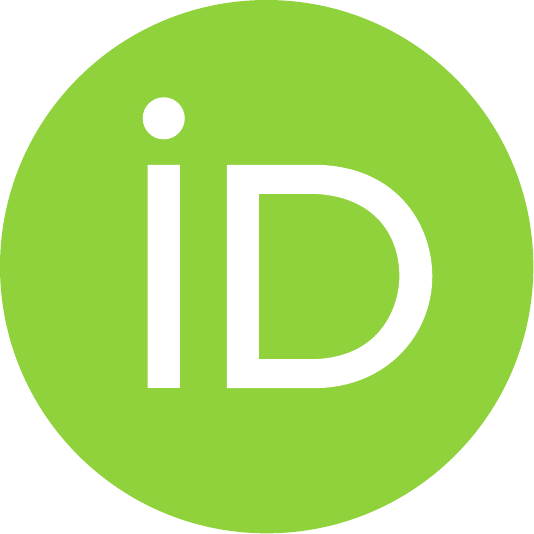}}}

\begin{document}

\ifthenelse{\equal{\anonymous}{no}}{

\title{Bichromatic Perfect Matchings with Crossings \thanks{O.A.\ and R.P.\ supported
		by the Austrian Science Fund (FWF): W1230.
		S.F.\ supported by the DFG Grant FE~340/13-1.
		M.S.\ supported by the DFG Grant SCHE~2214/1-1. We thank our anonymous reviewers for the useful suggestions.}}
\author{Oswin~Aichholzer\inst{1}\orcidID{0000-0002-2364-0583} \and
	Stefan~Felsner\inst{2}\orcidID{0000-0000-0000-0000} \and
	Rosna~Paul\inst{1}\orcidID{0000-0002-2458-6427} \and
	Manfred~Scheucher\inst{2}\orcidID{0000-0000-0000-0000} \and
	Birgit~Vogtenhuber\inst{1}\orcidID{0000-0002-7166-4467}
}
\authorrunning{O. Aichholzer, S. Felsner, R. Paul, M. Scheucher, B. Vogtenhuber}
\institute{Institute of Software Technology, Graz University of Technology, Graz, Austria\\
	\email{\{oaich,ropaul,bvogt,weinberger\}@ist.tugraz.at} \and
	Institute for Mathematics, Technical University of Berlin, Berlin, Germany\\
	\email{\{felsner,scheucher\}@math.tu-berlin.de} 
}

}{
\title{Bichromatic Perfect Matchings with Crossings}
\author{\textcolor{red}{anonymous author(s)}}
\authorrunning{anonymous}
}

\maketitle

\begin{abstract}
	We consider bichromatic point sets with $n$ red and $n$ blue points and study straight-line bichromatic perfect matchings on them. We show that every such point set in convex position admits a matching with at least $\frac{3n^2}{8}-\frac{n}{2}+c$ crossings, for some $ -\frac{1}{2} \leq c \leq \frac{1}{8}$. This bound is tight since for any $k> \frac{3n^2}{8} -\frac{n}{2}+\frac{1}{8}$ there exist bichromatic point sets that do not admit any perfect matching with $k$ crossings.
 
\end{abstract}

\section{Introduction}

Let $P = R \cup B$, $|R| = |B| = n$ be a point set in \emph{general position}, that is, no three points of $P$ are collinear. We refer to $R$ and $B$ as the set of red and blue points, respectively. A~straight-line matching $M$ on $P$ where every point in $R$ is uniquely matched to a point in $B$ is called a \emph{straight-line bichromatic perfect matching} (all matchings considered in this work are straight-line, so we will mostly omit this term).                                
In this work, we study the existence of bichromatic perfect matchings on~$P$ with a fixed number $k$ of crossings, where $0 \leq k \leq \binom{n}{2}$. 
It is folklore that any~$P$ of even size admits a crossing-free perfect matching.
Perfect matchings with $k$ crossings in the uncolored setting have been considered in~\cite{10.1007/978-3-031-06678-8_4}.
There it is shown that for every $k \leq \frac{n^2}{16} -O(n\sqrt{n})$, every point set of size $2n$ admits a perfect matching with exactly $k$ crossings and that there exist point sets where every perfect matching has at most $\frac{5n^2}{18}$ crossings. 
As a direct consequence, there exist bichromatic point sets which do not admit bichromatic perfect matchings with $k$ crossings for $k >\frac{5n^2}{18}$.
For $2n$ (uncolored) points in convex position it was shown in~\cite{10.1007/978-3-031-06678-8_4} that they
admit perfect matchings with $k$ crossings for every $k$ in the range from 0 to $\binom{n}{2}$.
\pagebreak 

For bichromatic point sets, this situation changes quite significantly. Consider a  
set $P$ of $2n$ points in convex position (convex point set, for short) with an alternating coloring, that is,
every second point along the convex hull is red (and the other points are blue). 
Moreover, let the number $n$ of red (and blue) points be even. 
Then  
the number of crossings in a bichromatic perfect matching~$M$ on~$P$ is at most $\frac{n(n-2)} {2} = \binom{n}{2} -\frac{n}{2}$. The idea is as follows: Label the points of~$P$ as $p_0,p_1, \ldots, p_{2n-1}$ along the boundary of the convex hull. 
Then $p_i$ cannot be matched to $p_{i+n}$ since both points have the same color. Hence, for any edge $e$ in~$M$, the number of crossings of~$e$ is at most $n-2$. 
As every crossing involves two edges, the number of crossings of~$M$
is at most $\frac{n(n-2)} {2} = \binom{n}{2} -\frac{n}{2}$.  
This bound is tight, since it is possible to construct a bichromatic perfect matching on $P$ with exactly $\binom{n}{2} -\frac{n}{2}$ crossings as follows. 
For $0 \leq i \leq n-1$,  match the point $p_i$ to the point $p_{i+n+1}$, when $i$ is even. Otherwise, match $p_i$  to $p_{i+n-1}$. \linebreak
Based on the above observations we state the following question.

\begin{ques}\label{op:1}
For which values of $k$ does
every bichromatic convex point set $P = R \cup B$, $|R| = |B| = n$,
admit a straight-line bichromatic perfect matching with exactly $k$ crossings?
\end{ques}

  The above example implies that if $k> \binom{n}{2}-\frac{n}{2}$, there exist bichromatic point sets with $n$ red and $n$ blue points that do not have any bichromatic perfect matching with $k$ crossings. Thus, the answer to Question~\ref{op:1} can be true only for $k \leq \binom{n}{2}-\frac{n}{2}$. As a main result of this paper, we prove the following theorem.

\begin{theorem} \label{thm:mainthm1}
	For every $n$ and for every $k > \frac{3n^2}{8} - \frac{n}{2} +\frac{1}{8}$, there exists a bichromatic convex point set 
	with $n$ red and $n$ blue points that does not have a straight-line bichromatic perfect matching with $k$ crossings.
\end{theorem}

To show this, we study bichromatic convex point sets and matchings on them with the maximum number of crossings. In Section~\ref{sec:bicroconv}, we first determine 
the maximum number of crossings for certain bichromatic convex point sets, depending  on their cardinality modulo $4$. Then we prove that this number gives a tight lower bound on the maximum number of crossings in any bichromatic convex point set. We further show some positive results for Question~\ref{op:1} in Section~\ref{sec:bichrogen}. \\

\noindent \textbf{Related work.}
A survey by Kano and Urrutia~\cite{ColoredPoints_Survey} gives an overview of various problems on bichromatic point sets, including matching problems.
Crossing-free bichromatic perfect matchings have been studied from various perspectives such as their structure~\cite{AsinowskiMR15,savic2022structural}, linear transformation distance~\cite{aichholzer2018linear}, and matchings compatible to each other~\cite{aichholzer2012compatible,aloupis2015bichromatic}. 
Sharir and Welzl~\cite{sharir2006number} proved that the  number of crossing-free bichromatic perfect matchings on $2n$ points is at most $O(7.61^{n})$.  However, on bichromatic perfect matchings with crossings much less is known. Pach et al.~\cite{prt_ppsdm_2021} showed that every straight-line drawing of $K_{n,n}$ contains a crossing family of size at least $n^{1-o(1)}$, where a \emph{crossing family} is a set of pairwise crossing edges. This implies that for any $P = R \cup B$, $|R| = |B| = n$, there exists a bichromatic perfect matching with at least $n^{2-o(1)}$ crossings.

 \section{Bichromatic Convex Point Sets}\label{sec:bicroconv}

Let $\mathcal{C}_{n,n}$ be the collection of all bichromatic convex point sets $P =  R \cup B$ with $|R| = |B| = n$.
For a point set $P \in \mathcal{C}_{n,n}$, we label the points in $P$ in clockwise direction along the convex hull as $p_0,p_1, \ldots, p_{2n-1}$ and refer to this as the \emph{clockwise ordering}. We will consider all indices modulo $2n$.
The number of crossings in a bichromatic perfect matching $M_P$ on $P$ is denoted by $\cro{M_P}$. If~$M_P$ has  
the maximum number of crossings among all such matchings on~$P$, then $M_P$ is called a \emph{max-crossing} matching on $P$. 
Among all max-crossing matchings for all $P \in \mathcal{C}_{n,n}$, we are interested in matchings
with the minimum number of crossings. 
We call such a matching a \emph{min-max-crossing matching} of~$\mathcal{C}_{n,n}$.  
From now on, we mostly refer to bichromatic perfect matchings just as \emph{matchings}.

A \emph{block} of $P \in \mathcal{C}_{n,n}$ is a maximal set of consecutive points of $P$ of the same color.
(If $R_1\!=\!\{p_{a}, p_{a+1}, \ldots, p_{a+s}\}$ is a red block then $p_{a-1}$ and $p_{a+s+1}$ are blue.)
Collecting the blocks of $P$ in clockwise order yields a cyclically ordered partition $(R_1,B_1, R_2, B_2, \ldots, R_s,B_s)$ of $P$. 
The coloring of $P$ is called 
\emph{$2s$--block} if it induces $s$ red and $s$ blue blocks.
In particular a \mbox{$2n$--block} coloring is alternating.
A \mbox{$2s$--block} coloring where all blocks have the same cardinality is
\emph{balanced}; see e.g.~Fig.~\ref{fig:cross}(a).
A balanced $2s$--block coloring only exists if $s$ divides~$n$. To overcome this restriction, we also consider $2s$--block colorings in which block sizes differ by at most 1 as \emph{balanced $2s$--block} colorings. Note that for $s=2$ and any given value of $n$, there is a unique balanced $2s$--block coloring (up to symmetry).

\begin{theorem} \label{thm:mainthm2}
	Let $P \in \mathcal{C}_{n,n}$ and $\M{P}$ be a max-crossing matching on $P$. Then
	\begin{equation*}
	\cro{\M{P}} \geq
	\begin{cases}
	\frac{3n^2}{8} - \frac{n}{2} & \text{if $n\equiv 0\mod{4}$}\\
	\frac{3n^2}{8} - \frac{n}{2} + \frac{1}{8} & \text{if $n\equiv 1\mod{4}$}\\
	\frac{3n^2}{8} - \frac{n}{2} - \frac{1}{2} & \text{if $n\equiv 2\mod{4}$}\\
	\frac{3n^2}{8} - \frac{n}{2} + \frac{1}{8} & \text{if $n\equiv 3\mod{4}$}
	\end{cases}       
	\end{equation*}
	Moreover, equality holds if $P$ has a balanced 4--block coloring. 
	In this case, $\M{P}$ is a min-max-crossing matching of~$\mathcal{C}_{n,n}$.
\end{theorem}

Theorem~\ref{thm:mainthm1} is implied by Theorem~\ref{thm:mainthm2}: consider any point set with balanced 4--block coloring.
The bound for the crossings in a max-crossing matching of this point set 
implies Theorem~\ref{thm:mainthm1}.
Theorem~\ref{thm:mainthm2} will follow directly from Lemma~\ref{lem:balancedexact} and Lemma~\ref{lem:partialsol2}, which are stated and shown in the next sections.

\subsection{Max-Crossing Matching of a Balanced  4--Block Coloring}

In this section, we determine
the number of crossings in any max-crossing matching of a point set with balanced 4--block coloring. 
A \emph{crossing family} of a point set $P$ is a set of edges spanned by points from $P$ that pairwise cross.

\begin{lemma}\label{lem:mexmat}
	Let $P \in \mathcal{C}_{n,n}$ have a 4--block coloring with blocks $R_1,B_1,R_2,B_2$ and let $\M{P}$ be a max-crossing matching on $P$. Then for each block $X \in  \{R_1,R_2,$ $B_1,B_2\}$, the edges emanating from $X$ form a crossing family. 
\end{lemma}

\begin{proof}
  Consider a block $X$ of $P$ 
  and assume w.l.o.g.~that $X = \{p_1,p_2, \dots, p_x\}$.
  If in a matching $M$ on $P$, there are two non-crossing edges with endpoints in $X$,
  then there also exist two such non-crossing edges with adjacent endpoints in $X$, 
  i.e., $M$ contains edges $(p_i,p_k)$ and $(p_{i+1},p_j)$ for some $1\leq i < x \leq j < k \leq 2n$.
  Let $M'$ be obtained from $M$ by replacing the two edges
  by the two crossing edges $(p_i,p_j)$ and $(p_{i+1},p_k)$ and note that $\cro{M'} = \cro{M} +1$.
\hfill\qed\end{proof}

\begin{figure}[htb]
	\centering
	\includegraphics[scale=0.5,page=1]{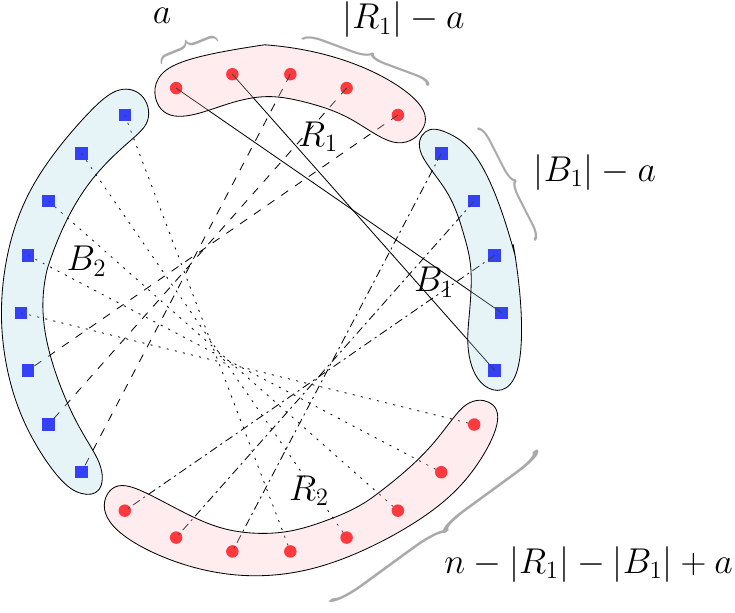}
	\caption{Structure of a max-crossing matching on a set $P$ with 4--block coloring.}
	\label{fig:4block}
\end{figure}

Consider a max-crossing matching $\M{P}$ on $P$. Lemma~\ref{lem:mexmat} implies that there
is some $a\geq 0$ such that in $\M{P}$, the first $a$ points of $R_1$ are
matched to points in $B_1$ while the last $|R_1|-a$ points of $R_1$ are
matched to points in $B_2$. 
Analogously, the first $|B_1|-a$ points of $B_1$ are matched to points in $R_2$.
Since $|R_2| = n- |R_1|$ and the $|B_1|-a$ points of $R_2$ are matched to $B_1$, 
the first $n - |R_1| - |B_1| + a$ points of $R_2$ are matched to
points in $B_2$; see Fig.~\ref{fig:4block}.
Hence, to get a max-crossing matching on $P$, it is sufficient to determine
the optimal value of~$a$.

By determining the optimum value of~$a$, we next construct a max-crossing matching on any set $P$ with balanced 4--block coloring and compute its exact crossing number.

\begin{restatable}{lemma}{lemmabalancedblock}\label{lem:balancedexact}
	Let $P \in \mathcal{C}_{n,n}$ have a balanced 4--block coloring and 
	let $\M{P}$ be a max-crossing matching on $P$. Then
	\begin{equation*}
	\cro{\M{P}} =
	\begin{cases}
	\frac{3n^2}{8} - \frac{n}{2} & \text{if $n\equiv 0\mod{4}$}\\
	\frac{3n^2}{8} - \frac{n}{2} + \frac{1}{8} & \text{if $n\equiv 1\mod{4}$}\\
	\frac{3n^2}{8} - \frac{n}{2} - \frac{1}{2} & \text{if $n\equiv 2\mod{4}$}\\
	\frac{3n^2}{8} - \frac{n}{2} + \frac{1}{8} & \text{if $n\equiv 3\mod{4}$}
	\end{cases}       
	\end{equation*}
\end{restatable}

\begin{proof}\label{lemm:balancedblock:proof}
	
	Let $P \in \mathcal{C}_{n,n}$ have a balanced 4--block coloring	with blocks $R_1,B_1,R_2$,$B_2$, 
	labeled such that $|R_1|,|B_1| \le \frac{n}{2}$. Let $r_1=|R_1|= \lfloor \frac{n}{2} \rfloor$ and $b_1=|B_1|= \lfloor \frac{n}{2} \rfloor$. Let~$M_P$ be a matching on $P$ such that the first $x$ points of the set $R_1$ 
	are matched to the last $x$ points of~$B_1$, as a crossing family, for some $x \in \mathbb{N}_0$. The number of pairs of non-crossing edges in $M_P$ is obtained by $(r_1-x)(b_1-x)+ x(n-r_1-b_1+x)= r_1b_1 - 2xb_1 - 2xr_1 +nx +2x^2$. 
	To determine the value $x \in \mathbb{N}_0$ that gives the maximum number of crossings, we first calculate the value $x^*\in\mathbb{R}$
	for which
	$f(x)= (n-2r_1-2b_1)x+2x^2+r_1b_1$ attains its minimum. This is achieved by $x^* = \frac{1}{2}(r_1+b_1-\frac{n}{2})$.
	Note that $x^*$ might not be in $\mathbb{N}_0$. Since $f$ is a quadratic function, its minimum over all $x \in \mathbb{N}_0$ is reached for
	$x=\lfloor x^* \rceil$, where $\lfloor x^* \rceil$ denotes the closest integer of $x^*\in\mathbb{R}$. Then the max-crossing matching $\M{P}$ on $P$
	has $\cro{\M{P}} = \binom{n}{2}-(n-2r_1-2b_1)x-2x^2-r_1b_1$ many crossings,
	where $x= \left\lfloor\frac{r_1+b_1}{2}-\frac{n}{4}\right\rceil = \left\lfloor\lfloor\frac{n}{2}\rfloor-\frac{n}{4}\right\rceil$.
	
	Note that the blocks in $P$ may differ in size by 1 and that also the rounding for obtaining $x$ depends on the value of $n \mod{4}$. 
	To account for this, we evaluate each case separately.
	
	\begin{description}
		\item [Case 1:] Let $n \equiv 0 \mod{4}$. Then there exists an integer $m$ such that $n = 4m$.\linebreak
		Then $r_1=b_1=2m$ and $x=\left\lfloor\frac{4m}{2}-\frac{4m}{4}\right\rceil=m$. 
		Hence the number of crossings of $\M{P}$ is 
		$\cro{\M{P}} = \binom{4m}{2}-(4m-4(2m))m-2m^2-(2m)^2 = 6m^2-2m$.
		Replacing $m$ by $\frac{n}{4}$ gives $\cro{\M{P}} = \frac{3n^2}{8} - \frac{n}{2}$.
		
		\item [Case 2:] Let $n \equiv 1 \mod{4}$. Then there exists an integer $m$ such that $n = 4m+1$.
		Then $r_1=b_1=2m$ and $x=\left\lfloor\frac{4m}{2}-\frac{4m+1}{4}\right\rceil=m$. 
		Hence the number of crossings of $\M{P}$ is 
		$\cro{\M{P}} = \binom{4m+1}{2}-(4m+1-4(2m))m-2m^2-(2m)^2 = 6 m^2 + m$. 
		Replacing $m$ by $\frac{n-1}{4}$ gives $\cro{\M{P}} = \frac{3n^2}{8} - \frac{n}{2} + \frac{1}{8}$.
		
		\item [Case 3:] Let $n \equiv 2 \mod{4}$. Then there exists an integer $m$ such that $n = 4m+2$.
		Then $r_1=b_1=2m+1$ and $x=\left\lfloor\frac{4m+2}{2}-\frac{4m+2}{4}\right\rceil=m$. 
		Hence the number of crossings of $\M{P}$ is 
		$\cro{\M{P}} = \binom{4m+2}{2}-(4m+2-4(2m+1))m-2m^2-(2m+1)^2 = 6m^2 + 4m.$
		Replacing $m$ by $\frac{n-2}{4}$ gives $\cro{\M{P}} = \frac{3n^2}{8} -\frac{n}{2} - \frac{1}{2}$.
		
		\item [Case 4:] Let $n \equiv 3 \mod{4}$. Then there exists an integer $m$ such that $n = 4m+3$.
		Then $r_1=b_1=2m+1$ and $x=\left\lfloor\frac{4m+2}{2}-\frac{4m+3}{4}\right\rceil=m$. 
		Hence the number of~crossings of $\M{P}$ is 
		$\cro{\M{P}} = \binom{4m+3}{2}-(4m+3-4(2m+1))m-2m^2-(2m+1)^2 = 6 m^2 + 7 m + 2$.
		Replacing $m$ by $\frac{n-2}{4}$ gives $\cro{\M{P}} = \frac{3n^2}{8} -\frac{n}{2} + \frac{1}{8}$.
	\end{description}

	\noindent Altogether this completes the proof of the lemma.
\hfill\qed\end{proof}

\subsection{Min-Max-Crossing Matching for All Colorings}

In the following, we show that the maximum number of crossings of a bichromatic matching on $P \in \mathcal{C}_{n,n}$
is minimized by sets with balanced 4--block coloring.
Let $P \in \mathcal{C}_{n,n}$. For any point $v \in P$, the point $w \in P$ is called the antipodal pair of $v$, if the line through $v$ and $w$ partitions $P$ into two equal sized halves (antipodals exist
because the number of points is even).  If antipodal pairs $v$ and $w$ are of the same color, then they are \emph{monochromatic antipodal} pairs (in short \emph{m-antipodal} pairs) and if they have different colors then they are \emph{bichromatic antipodal} pairs (in short \emph{b-antipodal} pairs).

\begin{restatable}{lemma} {lempartialsolb}\label{lem:partialsol2}
	Let $P,Q \in \mathcal{C}_{n,n}$, where $Q$ has a balanced 4--block coloring and let $\M{P}, \M{Q}$ be max-crossing matchings on $P$ and $Q$, respectively. Then $\cro{\M{P}} \geq \cro{\M{Q}}$. That is, $\M{Q}$ is a min-max-crossing matching of the set $\mathcal{C}_{n,n}$. 
\end{restatable}

To prove Lemma~\ref{lem:partialsol2}, we make use of a variant of the classic ham sandwich theorem~\cite{toth2017handbook}.
A full proof can be found in \Cref{app_lem3}.

\begin{proof}[sketch] 
	We define a matching $M_P$ on $P$ in three steps and then compare its crossings with those of $\M{Q}$, where we will distinguish two cases.
	 
\begin{description} 
	\item[Step 1:]
  Let $S$ be the point set obtained by removing all the b-antipodal pairs from $P$. If $S$ is empty then all the points in $P$ are b-antipodal pairs. Thus $P$ admits a crossing family of size $n$ (which is a perfect matching) that has more crossings than $\M{Q}$. Hence we may assume that $S$ is non-empty.

\item[Step 2:] Partition the set $S$ into four groups as follows. First, we arbitrarily partition $S$ into two consecutive sets $S_L$ and $S_R$ of equal size and note that each part contains half of the blue and half of the red points. Using the ham sandwich theorem, partition $S_L$ into $S_{L,1}$ and $S_{L,2}$ such that each of them has an equal number of red and blue points.
Due to the symmetry of $S_L$ and $S_R$, this partition can be duplicated on $S_{R,1}$. Depending on the ham sandwich cut, $S_{L,1}$, $S_{L,2}$, $S_{R,1}$, and $S_{R,2}$ form four or six bundles of consecutive  points along the convex hull. If we have only four bundles, we are done with the partition. So assume that we have six bundles. Then one partition in the part $S_R$, say $S_{R,2}$, is split into $S_{R,2a}$ and $S_{R,2b}$ by $S_{R,1}$ along the convex hull; see Fig.~\ref{fig:partition} (left). 
A similar splitting occurs for $S_L$.
 As the points in $S_{R,2b}$ and $S_{L,2a}$, and $S_{R,2a}$ and $S_{L,2b}$ are m-antipodal pairs, the composition of points in $S_{R,2}$ is same as in  $S_{L,2a} \cup S_{R,2a}$ and the composition of points in $S_{L,2}$ is same as  $S_{L,2b} \cup S_{R,2b}$. 
	That is, if we have four or six bundles, they can always be (re)assembled into
	four cyclically connected groups $S_{RT},S_{RB},S_{LB},S_{LT}$ 
	of $S$ such that each of these groups has the same number of
red and blue points; see again Fig.~\ref{fig:partition}.
 Moreover, each group is antipodal to another group: $S_{RT}$ is antipodal to $S_{LB}$ and $S_{LT}$ is antipodal to $S_{RB}$. We call these pairs of groups \emph{matching-pair} groups.

\begin{figure}[htb]
	\centering
	\includegraphics[scale=0.5,page=1]{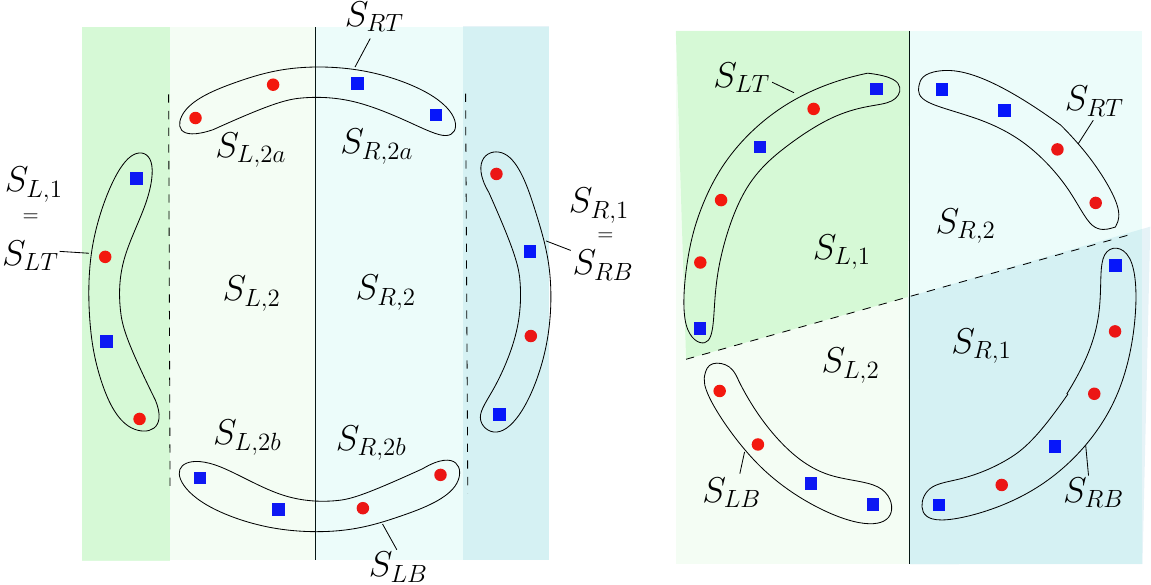}
	\caption{A bichromatic point set $S$ with 16 points (left) and 20 points (right). In both cases,
          dotted lines represent the partition of $S_L$ obtained by the ham sandwich theorem, and its mirror on $S_R$. }
	\label{fig:partition}
\end{figure}
		
\item[Step 3:] Add all the removed b-antipodal pairs back to $S$ to get $P$.
The partition of $S$ induces a partition $P_{RT},P_{RB},P_{LB},P_{LT}$ in
$P$. Note that the number of red (blue) points of $P_{RT}$ and the number of blue
(red) points of $P_{LB}$ are equal. The same holds for $P_{RB}$ and
$P_{LT}$. 	Define the matching $M_P$ on $P$ as follows. For any matching-pair group $(X,Y) \in \{(P_{RT}, P_{LB}), (P_{RB},P_{LT}) \}$, the points in $X$ are matched to points in $Y$ such that any two of the matching edges emanating from the points of the same color on $X$ cross each other.
\end{description}

Assuming that 
$S$ is non-empty, there are two possible structures for $S$, depending on whether $|S_{L}|/2$ is even or odd. We consider them as two separate cases; see~Fig.~\ref{fig:Scases}. Case 1: All groups in the partition of $S$ have the same number $m$ of red and blue points. Case 2: Each group in one matching-pair group has $m$ red (and blue)  points and each group in the other matching-pair has $m+1$ red (and blue) points.
	
	\begin{figure}[htb]
		\centering
		\includegraphics[scale=0.55,page=2]{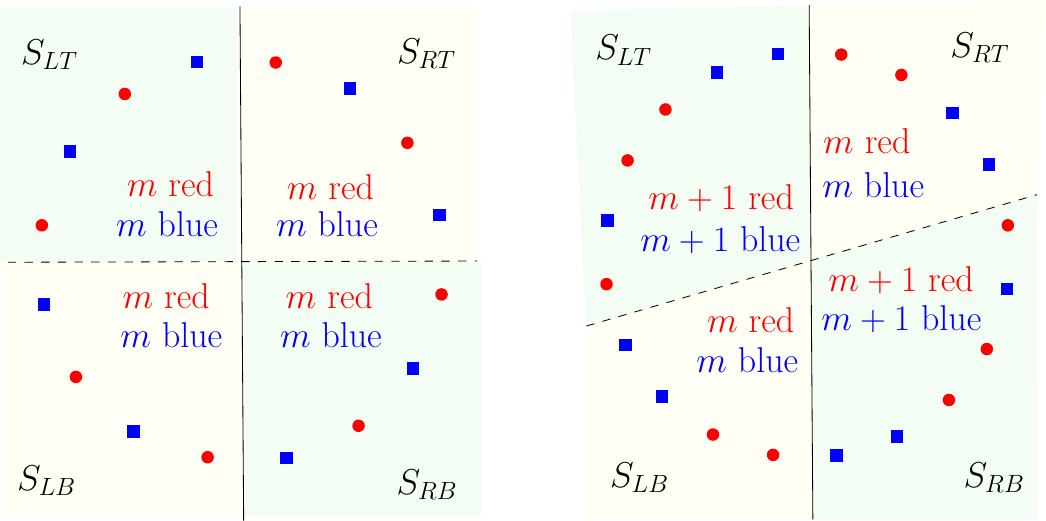}
		\caption{Possible distribution of red and blue points between the groups of $S$. }
		\label{fig:Scases}
	\end{figure}

	For Case 1,
  the number of crossings in $M_P$ is given by  \  
	$\cro{M_P} \geq \binom{m+x_1}{2}+ \binom{m+x_2}{2} +\binom{m+y_1}{2}+\binom{m+y_2}{2} +(2m+x_1+x_2)(2m+y_1+y_2)+x_1x_2+y_1y_2$, 
	where $x_1$, $x_2$  are the number of b-antipodal points in $P_{RT}$  of the colors red and blue, respectively and $y_1,y_2$ are the number of b-antipodal points in $P_{LT}$ of the colors red and blue. These b-antipodal pairs always cross in $M_P$ and contribute $x_1x_2 +y_1y_2$ crossings in $M_P$. 
	Then for a balanced 4--block coloring $Q \in \mathcal{C}_{n,n}$ we have, $\cro{\M{Q}} = \frac{3n^2}{8} - \frac{n}{2} +c$, where $n = 4m+x_1+x_2+y_1+y_2$ and $c \in \{ \frac{1}{8}, -\frac{1}{2} ,0\}$ by  Lemma~\ref{lem:balancedexact}. Comparing the number of crossings in $M_P$ and $\M{Q}$ gives \linebreak $
	\cro{M_P} - \cro{\M{Q}}  \geq \frac{(x_1y_1+x_2y_1+x_1y_2+x_2y_2)}{4} +\frac{(x_1^2+x_2^2)}{8}+\frac{(y_1^2+y_2^2)}{8}
	+ \frac{(x_1x_2+y_1y_2)}{4} -c $. As $c \leq \frac{1}{8}$ and since $S$ is a proper subset of~$P$, $x_1+x_2+y_1+y_2 \geq 1$. Thus $\cro{M_P} \geq \cro{\M{Q}}$.

The reasoning for Case 2 is similar to the one for Case 1 sketched above.
\hfill\qed\end{proof}

\begin{figure}[htb]
	\centering
	\includegraphics[scale=0.5,page=2]{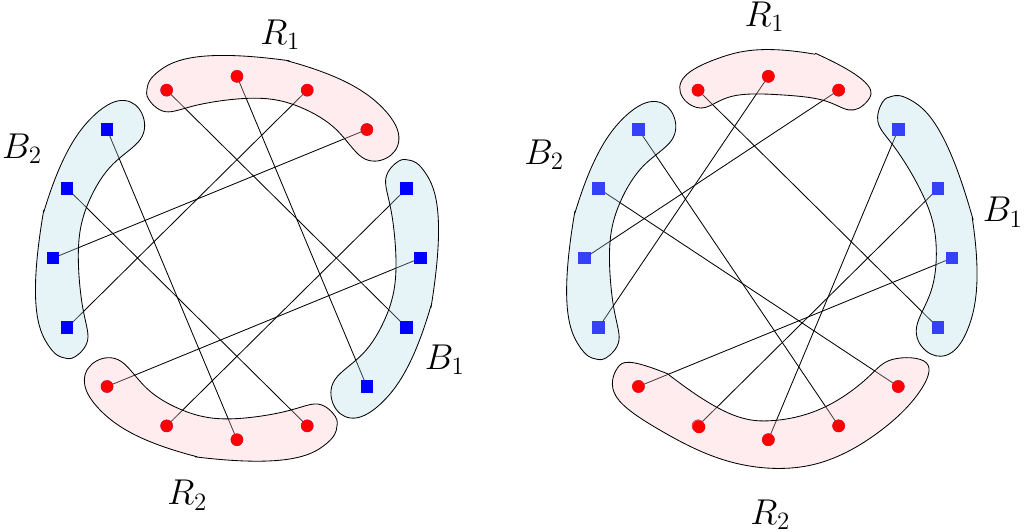}
	\caption{A balanced 4--block coloring (left) and a slightly unbalanced 4--block coloring (right) on 16 points and max-crossing matchings on them, each with 20 crossings.}
	\label{fig:cross}
\end{figure}

As mentioned, Theorem~\ref{thm:mainthm2} 
follows directly from Lemma~\ref{lem:balancedexact} and Lemma~\ref{lem:partialsol2}. \linebreak 
We remark that balanced 4--block colorings are not the only colorings that admit min-max-crossing matchings.
For example, for $n \equiv 0 \mod{4}$, a 4-block coloring with
block sizes $\frac{n}{2}+1,\frac{n}{2},\frac{n}{2}-1,\frac{n}{2}$ always induces the same number of crossings
as the according balanced 4--block coloring; 
see Fig.~\ref{fig:cross}.

\section{Further Results}\label{sec:bichrogen}
We have shown that for any $n$ and any $k > \frac{3n^2}{8}- \frac{n}{2} +\frac{1}{8}$ there exist point sets $P \in \mathcal{C}_{n,n}$ that do not admit any 
bichromatic perfect matching with $k$ crossings. It is natural to ask what happens in the range $[1,\frac{3n^2}{8}]$. By straight-forward calculations, 
any $P \in \mathcal{C}_{n,n}$ with $2n$--block coloring cannot have a bichromatic perfect matching with $k$ crossings for $k \in \{1,2\}$. 
Using computers (with SAT framework) 
we obtained that every $P \in \mathcal{C}_{7,7}$ admits bichromatic perfect matchings with $k$ crossings for any $k \in \{0, 1, \ldots, 15\}  \setminus\{1,2\}$.
Based on this, we can~show the following proposition. 
Its proof is deferred to \Cref{app_prop1}.

\begin{restatable}{proposition}{propoconvmore}\label{prop:convmore}
	For $n \geq 7$, every $P \in \mathcal{C}_{n,n}$  admits bichromatic perfect matchings with $k$ crossings for any $k \in \{0, 1, \ldots, \frac{15(n-6)}{7}\} \setminus\{1,2\}$.
\end{restatable}

For bichromatic point sets in general (non-convex) position with $n$ red and~$n$ blue points, computer assisted results shows that for $n=6$, every such set admits bichromatic perfect matchings with $k$ crossings for $k \in \{0,3,4\}$. 
With a similar proof as for Proposition~\ref{prop:convmore},
it follows that every bichromatic point set with $n$ red and $n$ blue points admits bichromatic perfect matchings with $k$ crossings for $k \in \{0, 1, \ldots, \frac{n}{3}\} \setminus \{1,2\}$. 
In ongoing work, we study 
the range $k \in [\frac{15(n-6)}{7},\frac{3n^2}{8}]$ for bichromatic convex point sets 
and the range $k \in [\frac{n}{3},\frac{5n^2}{18}]$ for bichromatic point sets in general position.

\section{Conclusion}

We considered max-crossing matchings of bichromatic convex point sets in the plane with $n$ red and $n$ blue points. 
We gave the exact number of crossings in max-crossing matchings of point sets with balanced 4--block coloring and showed that these matchings are min-max-crossing matchings of bichromatic convex point sets. 
This result implies a negative answer to Question~\ref{op:1} on the existence of matchings with $k$ crossings in the convex case for $k > \frac{3n^2}{8}- \frac{n}{2} +\frac{1}{8}$. We further answered the question for $k \le \frac{15(n-6)}{7}$. 
From a computational point of view, an interesting open question is the following: 
\begin{ques}
	Given a bichromatic (convex) point set $P$ and an integer $k$, what is the computational complexity of deciding whether 
	there is a matching with exactly \linebreak (or at least) $k$ crossings?
\end{ques}

\bibliographystyle{splncs04}
\bibliography{bichromatic}

\begin{thebibliography}{10}
\providecommand{\url}[1]{\texttt{#1}}
\providecommand{\urlprefix}{URL }
\providecommand{\doi}[1]{https://doi.org/#1}

\bibitem{aichholzer2018linear}
Aichholzer, O., Barba, L., Hackl, T., Pilz, A., Vogtenhuber, B.: Linear
  transformation distance for bichromatic matchings. Computational Geometry
  \textbf{68},  77--88 (2018). \doi{10.1016/j.comgeo.2017.05.003}

\bibitem{10.1007/978-3-031-06678-8_4}
Aichholzer, O., Fabila-Monroy, R., Kindermann, P., Parada, I., Paul, R., Perz,
  D., Schnider, P., Vogtenhuber, B.: Perfect matchings with crossings. In:
  Combinatorial Algorithms (IWOCA 2022). LNCS, vol. 13270, pp. 46--59. Springer
  (2022). \doi{10.1007/978-3-031-06678-8_4}

\bibitem{aichholzer2012compatible}
Aichholzer, O., Hurtado, F., Vogtenhuber, B.: Compatible matchings for
  bichromatic plane straight-line graphs. Proceedings of EuroCG'12 pp. 257--260
  (2012), \url{https://www.eurocg.org/2012/booklet.pdf}

\bibitem{aloupis2015bichromatic}
Aloupis, G., Barba, L., Langerman, S., Souvaine, D.L.: Bichromatic compatible
  matchings. Computational Geometry  \textbf{48}(8),  622--633 (2015).
  \doi{10.1016/j.comgeo.2014.08.009}

\bibitem{AsinowskiMR15}
Asinowski, A., Miltzow, T., Rote, G.: Quasi-parallel segments and
  characterization of unique bichromatic matchings. J. Comput. Geom.
  \textbf{6}(1),  185--219 (2015). \doi{10.20382/jocg.v6i1a8},
  \url{https://doi.org/10.20382/jocg.v6i1a8}

\bibitem{ColoredPoints_Survey}
Kano, M., Urrutia, J.: Discrete geometry on colored point sets in the plane —
  a survey. Graphs and Combinatorics  \textbf{37}(1),  1–53 (Jan 2021).
  \doi{10.1007/s00373-020-02210-8},
  \url{https://doi.org/10.1007/s00373-020-02210-8}

\bibitem{prt_ppsdm_2021}
Pach, J., Rubin, N., Tardos, G.: Planar point sets determine many pairwise
  crossing segments. Advances in Mathematics  \textbf{386},  107779 (2021).
  \doi{10.1016/j.aim.2021.107779}

\bibitem{savic2022structural}
Savi{\'c}, M., Stojakovi{\'c}, M.: Structural properties of bichromatic
  non-crossing matchings. Applied Mathematics and Computation  \textbf{415},
  126695 (2022)

\bibitem{sharir2006number}
Sharir, M., Welzl, E.: On the number of crossing-free matchings, cycles, and
  partitions. SIAM Journal on Computing  \textbf{36}(3),  695--720 (2006).
  \doi{10.1137/050636036}

\bibitem{toth2017handbook}
T\'oth, C.D., O'Rourke, J., Goodman, J.E.: {Handbook of Discrete and
  Computational Geometry}. CRC press, third edn. (2017).
  \doi{10.1201/9781315119601}

\end{thebibliography}

\appendix

\newpage

\section{Proof of Lemma 3}\label{app_lem3}

Before proving \Cref{lem:partialsol2}, we first show that the max-crossing matching of a balanced 4--block coloring is a min-max-crossing matching for the set of all $P \in \mathcal{C}_{n,n} $ with a 4--block coloring.

\begin{restatable}{lemma}{lempartialsola}\label{lem:partialsol1}
	Let $Q \in \mathcal{C}_{n,n}$ have a balanced 4--block coloring and 
	let $\M{Q}$ be a max-crossing matching on $Q$. Then $\M{Q}$ is a min-max-crossing matching for all 4--block colored point sets of size $2n$.
\end{restatable}

\begin{proof}\label{lemm:partialsola:proof}
	Let $P \in \mathcal{C}_{n,n}$ have a 4--block coloring	with blocks $R_1,B_1,R_2$,$B_2$, 
	labeled such that $|R_1|,|B_1| \le \frac{n}{2}$. Let $r_1=|R_1|$ and $b_1=|B_1|$.
  As mentioned in the proof of \Cref{lem:balancedexact}, the function $h(r_1,b_1)=(n-2r_1-2b_1)x+2x^2+r_1b_1$
  with $x=\max\{\left\lfloor\frac{1}{2}(r_1+b_1-\frac{n}{2})\right\rceil, 0\}$\footnote{Here we consider the maximum among $\left\lfloor\frac{1}{2}(r_1+b_1-\frac{n}{2})\right\rceil$ and zero to make sure that $x\geq 0$.} gives the number of non-crossing edge pairs in a max-crossing matching of $P$.
	To determine a point set with 4--block coloring such that its max-crossing matchings minimizes 
	the number of crossings among all max-crossing matchings of 4--block colored point sets, 
	we need to maximize  $h(r_1,b_1)$ over its domain, that is, over $r_1, b_1 \leq \frac{n}{2}$. 

	For that, we separately consider the cases when $x= \frac{1}{2}(r_1+b_1-\frac{n}{2})+q >0$, where $q$ is the smallest real number required to make $x$ an integer, and $x=0$.
	The value $q$ is the smallest real number required to make $x$ an integer. Note that~$q$ depends on the parity of $r_1 +b_1$ and on the value of $n \mod{4}$ (see Table~\ref{tab:tableforq}).
  
  \begin{table}[h!]
  	\begin{center}
		\begingroup
		\renewcommand{\arraystretch}{1.3}
  		\begin{tabular}{c|c|c|c|c} 
  			$r_1+b_1 \,$ & $ \, n \equiv 0 \mod 4 \,$ & $\, n \equiv 1 \mod 4 \,$ & $\, n \equiv 2 \mod 4\, $ & $\, n \equiv 3 \mod 4 \,$\\
  			\hline \hline
			even & 0 & $-\frac{1}{4}$ & $\frac{1}{2}$ & \ \ $\frac{1}{4}$\\[0.5ex]
  			\hline 
			odd & $\frac{1}{2}$ & \ \ $\frac{1}{4}$ & 0 & $-\frac{1}{4}$\\[0.5ex]
  		\end{tabular}
		\endgroup
  	\end{center}
  	\caption{Values of $q$ for Lemma~\ref{lem:partialsol1}.}
  	\label{tab:tableforq}
  \end{table}
  
  By substituting the value  $x = \frac{r_1+b_1}{2}-\frac{n}{4}+q$ in the function $h(r_1,b_1)$,  we obtain 
	$h(r_1,b_1) = \frac{nb_1+nr_1-b_1^2-r_1^2}{2}-\frac{n^2}{8} +2q^2$.
	The critical points of this function are obtained by setting its partial derivatives with respect to $r_1$ and $b_1$ to zero. This gives exactly one critical point at $(\frac{n}{2}, \frac{n}{2})$. The second partial derivatives show that this point is a local maximum of $h(r_1,b_1)$. But $\frac{n}{2}$ might be a real number or not have the correct parity for the requirement. Hence it might not be give valid maximum w.r.t.~$r_1,b_1,n$. 
	Thus, we have to separately consider the different parities of $r_1+b_1$ in combination with the different values of $n \mod 4$.
	For that, we consider the following cases. Note that as $h(r_1,b_1)$ is symmetric in $r_1$ and $b_1$, w.l.o.g. we can assume that $r_1 \leq b_1$.\vspace{2mm}
  
 \begin{description}
 	\item [Case 1:]  \textbf{If $n \equiv 0 \mod 4$:}
 	
 	    Then $n=4m$ for some integer $m$. In this case, the critical point $r_1 =b_1 = \frac{n}{2} = 2m$, is an integer. 
		In order to find a valid maximum, we also need to consider the parity of $r_1+b_1$. This leads to the following cases.
 	
 	   \begin{description}
 	   	\item[If $r_1+b_1$ is even:] 
 	   	 In this case,
 	   	$h(r_1,b_1) = \frac{nb_1+nr_1-b_1^2-r_1^2}{2}-\frac{n^2}{8}$. 
			   Then critical point $r_1=b_1= 2m$ gives the maximum of $h(r_1,b_1) = \frac{n^2}{8}$.
 	   	
 	   	\item [If $r_1+b_1$ is odd:] Here, the above critical point is not valid as the sum is not odd. In this case,
 	   	 $h(r_1,b_1) = \frac{nb_1+nr_1-b_1^2-r_1^2}{2}-\frac{n^2}{8} +\frac{1}{2}$. 
 	   	 The maximum is at $r_1 = 2m$ and $b_1 = 2m+1$ and its value is $h(r_1,b_1) = \frac{n^2}{8}$.
 	   \end{description}
    
    	\item [Case 2:]  \textbf{If $n \equiv 1 \mod 4$:}
    
    Then $n=4m+1$ for some integer $m$. In this case the critical point, $r_1 =b_1 = \frac{n}{2}$, does not give an integer value. 
		 In order to find the maximum, we also need to consider the parity of $r_1+b_1$. This leads to the following cases.
    
    \begin{description}
    	\item[If $r_1+b_1$ is even:] 
    	In this case,
    	$h(r_1,b_1) = \frac{nb_1+nr_1-b_1^2-r_1^2}{2}-\frac{n^2}{8} +\frac{1}{8}$. 
			Then the maximum is at $r_1=b_1 = 2m$ and has the value $h(r_1,b_1) = \frac{n^2}{8} - \frac{1}{8}$.
    	
    	\item [If $r_1+b_1$ is odd:] In this case,
    	$h(r_1,b_1) = \frac{nb_1+nr_1-b_1^2-r_1^2}{2}-\frac{n^2}{8} +\frac{1}{8}$. 
    	The maximum is obtained at $r_1 = 2m$ and $b_1 = 2m+1$ and has the value $h(r_1,b_1) = \frac{n^2}{8}- \frac{1}{8}$.
    \end{description}

	\item [Case 3:]  \textbf{If $n \equiv 2 \mod 4$:}

Then $n=4m+2$ for some integer $m$. In this case the critical point, $r_1 =b_1 = \frac{n}{2}$, is an integer value. 
		 However, for the maximum, we also need to consider the parity of $r_1+b_1$. This leads to the following cases.

\begin{description}
	\item[If $r_1+b_1$ is even:] 
	In this case,
	$h(r_1,b_1) = \frac{nb_1+nr_1-b_1^2-r_1^2}{2}-\frac{n^2}{8} +\frac{1}{2}$. 
		Then $r_1=b_1 = 2m+1$ gives a valid point and the resulting maximum value is $h(r_1,b_1) = \frac{n^2}{8} + \frac{1}{2}$.
	
	\item [If $r_1+b_1$ is odd:] In this case,
	$h(r_1,b_1) = \frac{nb_1+nr_1-b_1^2-r_1^2}{2}-\frac{n^2}{8}$. 
	The maximum is obtained at $r_1 = 2m$ and $b_1 = 2m+1$ and has the value $h(r_1,b_1) = \frac{n^2}{8}- \frac{1}{2}$.
\end{description}

	\item [Case 4:]  \textbf{If $n \equiv 3 \mod 4$:}

Then $n=4m+3$ for some integer $m$. In this case the critical point, $r_1 =b_1 = \frac{n}{2}$, does not have integer values. 
		 In order to find the points reaching the maximum, we again also consider the parity of $r_1+b_1$. This leads to the following cases.

\begin{description}
	\item[If $r_1+b_1$ is even:] 
	In this case,
	$h(r_1,b_1) = \frac{nb_1+nr_1-b_1^2-r_1^2}{2}-\frac{n^2}{8} +\frac{1}{8}$. 
		Then maximum is at $r_1=b_1 = 2m+2$ and has the value $h(r_1,b_1) = \frac{n^2}{8} - \frac{1}{8}$.
	
	\item [If $r_1+b_1$ is odd:] In this case,
	$h(r_1,b_1) = \frac{nb_1+nr_1-b_1^2-r_1^2}{2}-\frac{n^2}{8} + \frac{1}{8}$. 
	Then maximum is obtained at $r_1 = 2m+1$ and $b_1 = 2m+2$ and has the value $h(r_1,b_1) = \frac{n^2}{8}- \frac{1}{8}$.
\end{description}
\end{description}

 In all the above cases, any point set $P$ with balanced 4--block coloring gives a min-max-crossing matching.

 \medskip
The only remaining case is to check what happens if when $x=0$. Here, $h(r_1,b_1) = r_1b_1$. 
We have to maximize this function under the restriction that $r_1+b_1 <\frac{n}{2}$ and $r_1, b_1 \geq 0$. 
Thus we get $ r_1 = b_1 = \frac{n}{4}$, which again might not be an integer or not have the right parity of $r_1+b_1$.
However, doing an analogous case distinction as above, the maximum number of non-crossing edges always is less than and equal to $ m^2 + am$, 
where $a$ is a fixed integer depending on the case.
This is relevantly smaller than the number of non-crossing edges in the maxima obtained above (by the balanced 4--block coloring),
which is given by $2m^2 + cm+d$, for some integers $c$ and $d$.
Hence no 4--block colored point set $P$ with $ r_1 + b_1 \leq \frac{n}{2}$ minimizes the number of non-crossing edges in 
$\M{P}$ over all point sets with 4--block coloring. 
 Altogether, this implies that the number of crossings in a max-crossing matching is minimized on a balanced 4--block coloring. 
\hfill\qed\end{proof}

\lempartialsolb*

\begin{proof}\label{lem:partialsolb:proof}
  We define a matching $M_P$ on $P$ in three steps and then compare its crossings with those of $\M{Q}$, where we distinguish two cases.
	\begin{description} 
		\item[Step 1:] 
			Let $S$ be the subset of $P$ obtained by removing all the b-antipodal pairs from $P$. Then $S$ is also a bichromatic convex point set with an equal number of red and blue points. Since each point in $S$ has an antipodal pair of the same color, the number of red (blue) points must be even. 
		
		If $S$ is empty then all the points in $P$ are b-antipodal pairs. Thus $P$ admits a crossing family of size $n$. Then clearly this crossing family (which is a perfect matching) has more crossings than $\M{Q}$.
		
	\item[Step 2:] Partition the set $S$ into four groups $S_{RT}, S_{RB},S_{LB},S_{LT}$ as follows. First, we partition the points of $S$ into two sets $S_L$ and $S_R$ of equal size. Note that each of these sets has an equal number of red and blue points. Using the ham sandwich theorem, we partition $S_L$ into $S_{L,1}$ and $S_{L,2}$ such that $S_{L,1}$ has the same number of red and blue points, and also $S_{L,2}$ has the same number of red and blue points. Note that if the number of red (blue) points in $S_L$ is odd, then the number on red (blue) points in $S_{L,1}$ and $S_{L,2}$ will differ by one. Due to the symmetry of $S_L$ and $S_R$ we have a corresponding partition  $S_{R,1}$ and $S_{R,2}$ of $S_R$. Depending on the ham sandwich cut, $S_{L,1}$, $S_{L,2}$, $S_{R,1}$, and $S_{R,2}$  form four or six bundles of consecutive  points along the convex hull. If we have only four bundles, we are done with the partition. So assume that we have six bundles. Then one partition in the part $S_R$, say $S_{R,2}$, is split into $S_{R,2a}$ and $S_{R,2b}$ by $S_{R,1}$ along the convex hull; see Fig.~\ref{fig:partition} (left). 
	A similar splitting occurs for $S_L$.
	As the points in $S_{R,2b}$ and $S_{L,2a}$, and $S_{R,2a}$ and $S_{L,2b}$ are m-antipodal pairs, the composition of points in $S_{R,2}$ is same as in  $S_{L,2a} \cup S_{R,2a}$ and the composition of points in $S_{L,2}$ is same as  $S_{L,2b} \cup S_{R,2b}$. 
	That is, if we have four or six bundles, they can always be (re)assembled into
	four cyclically connected groups $S_{RT},S_{RB},S_{LB},S_{LT}$ 
	of $S$ such that each of these groups has the same number of
	red and blue points; see again Fig.~\ref{fig:partition}. 
	Moreover, each group is antipodal to another group: $S_{RT}$ is antipodal to $S_{LB}$ and $S_{LT}$ is antipodal to $S_{RB}$. We call these pairs of groups \emph{matching-pair} groups.

	 If $S =P$, sort the points in $S_{RT}$ and $S_{LB}$ such that all the red points appear before the blue points w.r.t.\ clockwise order. Then sort the points in $S_{RB}$ and $S_{LT}$ such that all the blue points appear before the red points. This gives a bichromatic point set $W$ with the partition $W_{RT},W_{RB},W_{LB},W_{LT}$.
		 
		 Define the matchings $M_S$ and $M_W$ on $S$ and $W$, respectively, as follows. For any $(X,Y) \in \{(S_{RT}, S_{LB}), (S_{RB},S_{LT}),(W_{RT}, W_{LB}), (W_{RB},W_{LT}) \}$, the points in $X$ are matched to points in $Y$ such that any two of the matching edges emanating from the same colored points on $X$ cross each other. Hence the matching edges of $X$ give two crossing families, where the size of each family is determined by the number of points in $X$ of each color. By our construction, the size of the crossing families is the same in both $S$ and~$W$. But in $S$ these crossing families might cross each other, while in $W$ they do not cross each other. Hence, $\cro{M_S} \geq \cro{M_W}$. By construction~$W$ has a balanced 4--block coloring (i.e., $W=Q$) and $M_W = \M{Q}$. Thus,
		 $\cro{M_S} \geq \cro{\M{Q}}$. However, the constructed matching $M_S$ might not be a max-crossing matching on $S$. Hence, if $\M{S}$ is a max-crossing matching for~$P$, then $\cro{\M{S}} \geq \cro{M_S}$. This implies $\cro{\M{S}} \geq \cro{\M{Q}}$, which completes the proof for the case $S=P$.
		 
		 For the remaining part, $S$ is a non-empty proper subset of $P$.
		
	 \item[Step 3:] Add back the b-antipodal pairs to $S$ to obtain $P$ and assign each b-antipodal pairs to one of the matching-pair groups. We denote the new groups by $P_{RT},P_{RB},P_{LB},P_{LT}$, respectively. This gives a partition of $P$ into four groups, where each group may have a different number of red and blue points, but the number of red (blue) points in one group is equal to the number of blue (red) points in its matching-pair group. For calculation purposes we assume that there are $x_1$  new red points and $x_2$ new blue points added to the group $P_{RT}$ and $y_1$  new red points and $y_2$ new blue points added to the group $P_{LT}$.
		
	\end{description}
	
 As we assume that $S$ is non-empty, there are two possible structures for $S$ as described in Step 2. We consider them as two separate cases. Case 1: All groups in the matching-pair groups have the same number $m$ of red and blue points. Case 2: Each group in one matching-pair group has $m$ red (blue)  points whereas each group in the other matching-pair has $m+1$ red (blue) points, see Fig.~\ref{fig:Scases}.

	\begin{description}
		\item[Case 1:] \textbf{All groups have $m$ red and $m$ blue points}

		\begin{figure}[htb]
			\centering
			\includegraphics[scale=0.5,page=3]{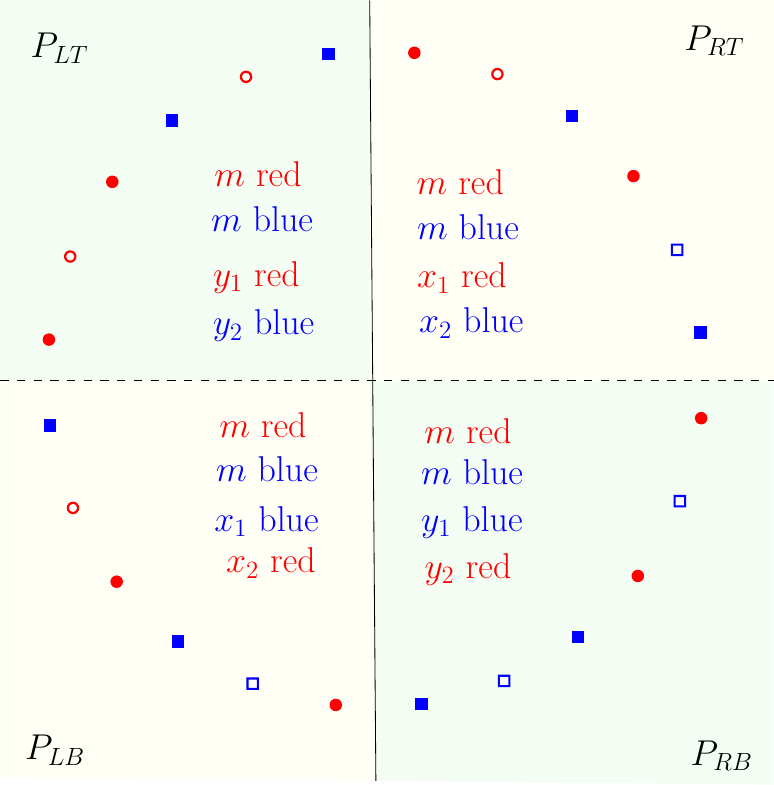}
			\caption{Distribution of points in Case 1 of the proof of Lemma~\ref{lem:partialsol2}. Filled vertices represent m-antipodal pairs and unfilled vertices represent b-antipodal pairs. }
			\label{fig:Pcase1}
		\end{figure}
		
		Define the matching $M_P$ on $P$ as follows. For any matching-pair group $(X,Y) \in \{(P_{RT}, P_{LB}), (P_{RB},P_{LT}) \}$, the points in $X$ are matched to points in $Y$ such that any two of the matching edges emanating from the same colored points in $X$ cross each other. Since $P_{RT}$ has $m+x_1$ red points, the edges emanating from these points form a crossing family of size $m+x_1$ with $\binom{m+x_1}{2}$ crossings in $M_P$. Similarly, the edges emanating from the $m+x_2$ blue points in $P_{RT}$ form a crossing family of size $m+x_2$ with $\binom{m+x_2}{2}$ crossings in~$M_P$. Note that these crossing families can cross each other in $M_P$. Likewise, $P_{LT}$ admits a crossing family of red points size $m+y_1$ with $\binom{m+y_1}{2}$ crossings and a crossing family of blue points of size $m+y_2$ with $\binom{m+y_2}{2}$ crossings. Furthermore, all the edges emanating from $P_{RT}$ cross all the edges emanating from $P_{LT}$, see Fig.~\ref{fig:Pcase1}.

		The number of crossings in $M_P$ is given by 
		\begin{align*}
			\cro{M_P} \geq & \, \binom{m+x_1}{2}+ \binom{m+x_2}{2} +\binom{m+y_1}{2}+\binom{m+y_2}{2}\\
		&+(2m+x_1+x_2)(2m+y_1+y_2)+a+b.
		\end{align*}

	    Here, $a$ is the number of crossings obtained between the bundle of edges emanating from the $x_1$ red points in $P_{RT}$ with the edges emanating from the $x_2$ blue points in $P_{RT}$. Similarly,  $b$ is the number of crossings between the edges emanating from the b-antipodal points in $P_{LT}$ of different color. Note that in this counting we did not count the crossings between the edges emanating from the red and blue points in $S_{RT}$ and $S_{LT}$. Also, the crossings between the edges emanating from the $x_1$ (or $x_2$) newly added red (blue) points in $P_{RT}$ and $m$ blue (red) points in the corresponding $S_{RT}$ were not considered in the above counting. Similarly for $P_{LT}$. 
		
		\begin{claim}
			$a \geq x_1\cdot x_2$ and $b \geq y_1\cdot y_2$.
			\end{claim}

		\begin{claimproof}
			In the following we prove $a = x_1\cdot x_2$. The proof for $b = y_1\cdot y_2$ follows similarly. Assume that $\{(p_1,q_1), (p_2,q_2)\}$ with
			$p_1,p_2 \in P_{RT}$ and $q_1,q_2 \in P_{LB}$ are the b-antipodal pairs such that $p_1,q_2$ are red and $p_2,q_1$ are blue (see Fig.~\ref{fig:redblucross}). Let $e_{v}$ be the matching edge of $M_P$ incident to point~$v$. We will show that either $e_{p_1}$ or $e_{q_1}$ crosses with $e_{p_2}$ or $e_{q_2}$.
			
				\begin{figure}[htb]
				\centering
				\includegraphics[scale=0.5,page=5]{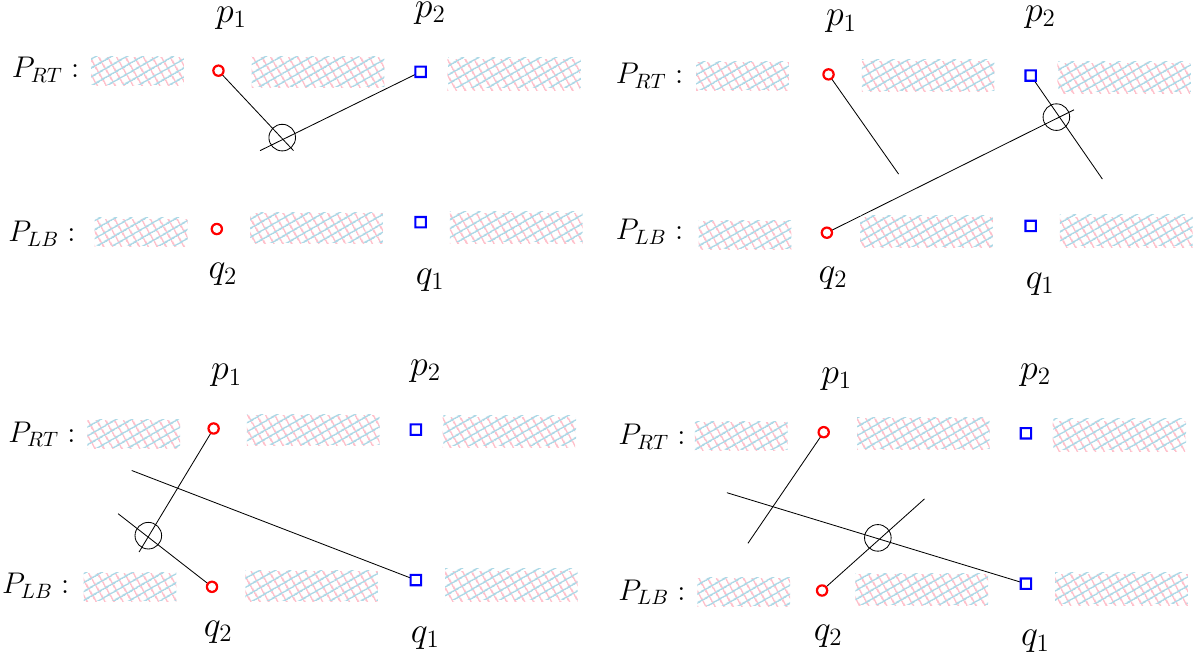}
				\caption{ Possible ways in which b-antipodal points cross in $M_P$. The crossings counted in the claim are marked by a circle. The crossed red-blue lines represents the remaining red-blue points in $P_{RT}$ and $P_{LB}$.}
				\label{fig:redblucross}
			\end{figure}
			
			Note that the edges $e_{p_1}$ and $e_{q_1}$ cross each other as they belong to the same crossing family and similarly, $e_{p_2}$ and $e_{q_2}$ cross each other. If the edge $e_{p_1}$ separates the points $p_2$ and $q_2$, then either $e_{p_2}$ or $e_{q_2}$ has to cross $e_{p_1}$ as $e_{p_2}$ and $e_{q_2}$ cross each other. If the edge $e_{p_1}$ does not separate the points $p_2$ and~$q_2$, then $e_{q_1}$ separates them as $e_{p_1}$ and $e_{q_1}$ cross each other. Thus, either $e_{p_1}$ or $e_{q_1}$ has to cross $e_{q_2}$. 
		\end{claimproof}
		\medskip	
          		
		By substituting the values of $a$ and $b$ we get
		\begin{align*}
			\cro{M_P} \geq & \, 6m^2 + 3m(x_1+x_2+y_1+y_2) -2m + x_1y_1+x_1y_2+x_2y_1+x_2y_2\\
		&+\frac{x_2^2+x_1^2+y_1^2+y_2^2-x_1-x_2-y_1-y_2}{2} + x_1x_2 + y_1y_2.
		\end{align*}

	    Consider a balanced 4--block coloring $Q \in \mathcal{C}_{n,n}$ with $n = 4m+x_1+x_2+y_1+y_2$. Then by Lemma~\ref{lem:balancedexact}, we get
	   \begin{align*}
	   \cro{\M{Q}} &= \frac{3n^2}{8} - \frac{n}{2} +c\\
	   &= \frac{3(4m+x_1+x_2+y_1+y_2)^2}{8} - \frac{4m+x_1+x_2+y_1+y_2}{2} +c\\
	   &= 6m^2 + 3m(x_1+x_2+y_1+y_2) +\frac{3(x_1y_1+x_2y_1+x_1y_2+x_2y_2)}{4}\\
	   &\quad \, + \frac{3(x_1x_2+y_1y_2)}{4}+ \frac{3(x_1^2+x_2^2)}{8}+\frac{3(y_1^2+y_2^2)}{8} -2m -\frac{x_1}{2} -\frac{x_2}{2}\\
	   &\quad \, -\frac{y_1}{2}-\frac{y_2}{2} +c,
	   \end{align*}
	   where $c$ takes one of the values $\frac{1}{8}, -\frac{1}{2}$,or $0$, depending on the divisibility of~$n$ by four.
	   Comparing the number of crossings in $M_P$ and $\M{Q}$ gives
	   \begin{align*}
		   \cro{M_P} - \cro{\M{Q}}  \geq & \, \frac{(x_1y_1+x_2y_1+x_1y_2+x_2y_2)}{4} +\frac{(x_1^2+x_2^2)}{8}+\frac{(y_1^2+y_2^2)}{8}\\
	   &+ \frac{(x_1x_2+y_1y_2)}{4} -c.
	   \end{align*}
	  
	  As $c \leq \frac{1}{8}$ and since $S$ is a proper subset of $P$ at least one among $x_1,x_2,y_1,y_2$ is at least 1. Thus $\cro{M_P} \geq \cro{\M{Q}}$.\\

	\item[Case 2:] \textbf{ Two groups in a matching-pair group have $m$ red (blue) points and the two remaining groups have $m+1$ red (blue) points.}

		As in Case 1, define the matching $M_P$ on $P$ as follows. For any matching-pair group $(X,Y) \in \{(P_{RT}, P_{LB}), (P_{RB},P_{LT}) \}$, the points in $X$ are matched to points in $Y$ such that any two of the matching edges emanating from the same colored points on $X$ cross each other. Since $P_{RT}$ has $m+x_1$ red points, the edges emanating from these points form a crossing family of size $m+x_1$ with $\binom{m+x_1}{2}$ crossings in~$M_P$. Similarly, the edges emanating from the $m+x_2$ blue points in $P_{RT}$ form a crossing family of size $m+x_2$ with $\binom{m+x_2}{2}$ crossings in~$M_P$. Note that these crossing families can cross each other in~$M_P$. Likewise, $P_{LT}$ admits a crossing family emanating from red points of size $m+y_1+1$ with $\binom{m+y_1+1}{2}$ crossings and a crossing family emanating from blue points of size $m+y_2+1$ with $\binom{m+y_2+1}{2}$ crossings. Furthermore, all the edges emanating from $P_{RT}$ cross all the edges emanating from $P_{LT}$, see Fig.~\ref{fig:Pcase2}. 

		\begin{figure}[htb]
			\centering
			\includegraphics[scale=0.5,page=4]{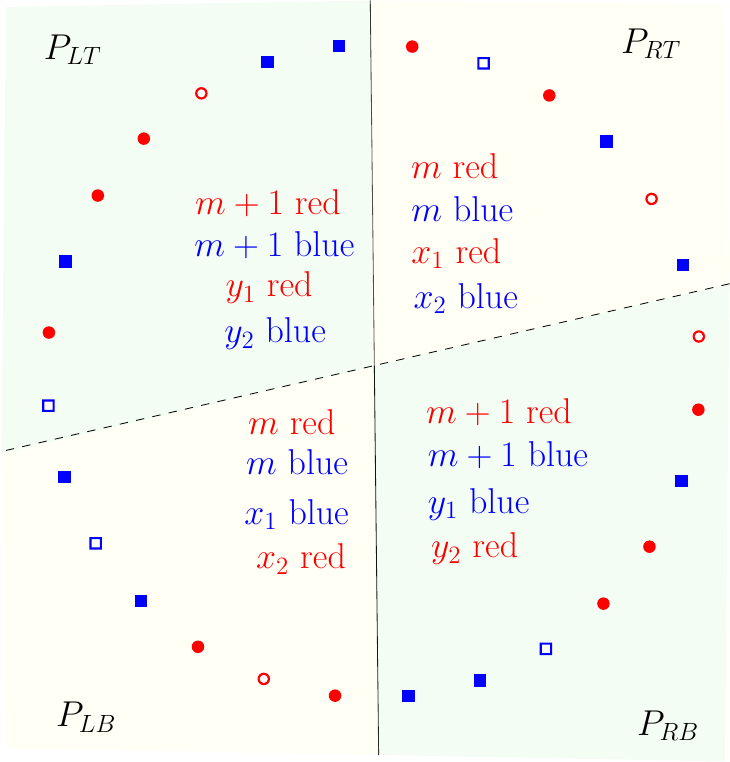}
			\caption{Distribution of points in Case 2 of the proof of Lemma~\ref{lem:partialsol2}.  
				Filled vertices represents m-antipodal pairs and unfilled vertices represents b-antipodal pairs.}
			\label{fig:Pcase2}
		\end{figure}

		The number of crossings in $M_P$ is given by 
		\begin{align*}
			\cro{M_P} \geq & \, \binom{m+x_1}{2}+ \binom{m+x_2}{2} +\binom{m+y_1+1}{2}+\binom{m+y_2+1}{2}\\
			&  + (2m+x_1+x_2)(2m+y_1+y_2+2)+a+b.
		\end{align*}

		 Here $a$ is the number of crossings obtained between the bundle of edges emanating from the $x_1$ red points in $P_{RT}$ with the edges emanating from the $x_2$ blue points in $P_{RT}$. Similarly, $b$ is the number of crossings between the edges emanating from the b-antipodal points in $P_{LT}$. Note that in this counting we did not count the crossings between the edges emanating from the red and blue points in $S_{RT}$ and $S_{LT}$. Also, the crossings between the edges emanating from the $x_1$ (or $x_2$) newly added red (blue) points in $P_{RT}$ and $m$ blue (red) points in the corresponding $S_{RT}$ were not considered in the above counting. Similarly for $P_{LT}$.  Analogous Case 1, we have $a \geq x_1x_2$ and $b \geq y_1y_2$. By substituting the values of $a$ and $b$ we get
		\begin{align*}
			\cro{M_P} \geq & \, 6m^2 + 3m(x_1+x_2+y_1+y_2) +4m + x_1y_1+x_1y_2+x_2y_1+x_2y_2\\
		&+\frac{x_2^2+x_1^2+y_1^2+y_2^2}{2}+ \frac{3(x_1+x_2)}{2}+ \frac{(y_1+y_2)}{2}+ x_1x_2 + y_1y_2.
		\end{align*}

    Consider a balanced 4--block coloring $Q \in \mathcal{C}_{n,n}$ with $n = 4m+2+x_1+x_2+y_1+y_2$. Then by Lemma~\ref{lem:balancedexact}, we get
	
	\begin{align*}
	\cro{\M{Q}} &= \frac{3n^2}{8} - \frac{n}{2} +c\\
	&= \frac{3(4m+x_1+x_2+y_1+y_2+2)^2}{8} - \frac{4m+x_1+x_2+y_1+y_2+2}{2} +c\\
	&= 6m^2 + 3m(x_1+x_2+y_1+y_2) +\frac{3(x_1y_1+x_2y_1+x_1y_2+x_2y_2)}{4}\\
	&\quad \, + \frac{3(x_1x_2+y_1y_2)}{4}+ \frac{3(x_1^2+x_2^2)}{8}+\frac{3(y_1^2+y_2^2)}{8} +4m +x_1 +x_2\\
	&\quad \, +y_1+y_2 +\frac{1}{2}+c,
	\end{align*}
	
	where $c$ takes one of the values $\frac{1}{8}, -\frac{1}{2}$,or $0$, depending on the divisibility of~$n$ by four.
	Comparing the number of crossings in $M_P$ and $\M{Q}$ gives
	
	 \begin{align*}
		 \cro{M_P} - \cro{\M{Q}} \geq & \, \frac{(x_1y_1+x_2y_1+x_1y_2+x_2y_2)}{4} +\frac{(x_1^2+x_2^2)}{8}+\frac{(y_1^2+y_2^2)}{8}\\
	&+ \frac{(x_1x_2+y_1y_2)}{4} + \frac{x_1+x_2}{2}-\frac{y_1+y_2}{2} -\frac{1}{2} -c.
	\end{align*}

  For $x_1+x_2 \geq 2$,  we have	$\cro{M_P} - \cro{\M{Q}} \geq \frac{(y_1+y_2)^2}{8} + \frac{1}{2} -c$. As $c \leq \frac{1}{8}$, $\cro{M_P} \geq \cro{\M{Q}}$. Thus we only have to consider the case when $x_1 + x_2 <2$, i.e., either $x_1 = x_2 =0$ or $x_1 = 1$ and $x_2 =0$. Note that we do not have to consider the case where $x_1 =0$ and $x_2 =1$ separately, as the inequality is symmetric in $x_1$ and $x_2$ (as well as in $y_1$ and $y_2$).

 	\item[Case 2.1:] Assume that $x_1=1$ and $x_2 =0$. It implies that $\cro{M_P} - \cro{\M{Q}} \geq \frac{(y_1 +y_2)(y_1 +y_2 -2)}{8} + \frac{1}{8} -c \geq \frac{(y_1 +y_2)(y_1 +y_2 -2)}{8}$, where the last inequality holds as $c \leq \frac{1}{8}$. Thus, $\cro{M_P} \geq \cro{\M{Q}}$ whenever $y_1+y_2 \neq 1$. W.l.o.g. assume that $y_1 =1$ and $y_2 =0$. Then the total number of red points in such a point set $P$ is $ n= 4m+4$. As $n \equiv 0 \mod{4}$, $c=0$ by Lemma~\ref{lem:balancedexact}. This implies that $\cro{M_P} \geq \cro{\M{Q}}$. 
 	
 	\item[Case 2.2:] Assume that $x_1 = x_2 = 0$. Then $\cro{M_P} - \cro{\M{Q}} \geq \frac{(y_1 +y_2)(y_1 +y_2 -4)}{8} - \frac{1}{2} -c \geq \frac{(y_1 +y_2)(y_1 +y_2 -4)}{8} -\frac{1}{2} - \frac{1}{8}$, where the last inequality holds as $c \leq \frac{1}{8}$. Thus, $\cro{M_P} \geq \cro{\M{Q}}$ whenever $y_1+y_2 \geq 5$. Assume that $y_1+y_2 = 4$, then the total number of red points in such a set $P$ is $ n= 4m+6$, i.e., $c= -\frac{1}{2}$. Thus, $\cro{M_P} \geq \cro{\M{Q}}$. Next, assume that $y_1+y_2 =1$. Then the total number of red points in such a set $P$ is $ n= 4m+3$, i.e., $c= \frac{1}{8}$. Thus, $\cro{M_P} - \cro{\M{Q}} \geq -1$. Similarly, assume that $y_1+y_2 =2$. Then the total number of red points in such a set $P$ is $ n= 4m+4$, i.e., $c= 0$. Thus, $\cro{M_P} - \cro{\M{Q}} \geq -1$.
 	And finally if $y_1+y_2 =3$. Then the total number of red points in such a set $P$ is $ n= 4m+5$, i.e., $c= \frac{1}{8}$. Thus, $\cro{M_P} - \cro{\M{Q}} \geq -1$. Note that $y_1+y_2 =0$ would imply $x_1+x_2+y_1+y_2 =0$ which imply $S=P$ which we considered already.
 	
 	  In short, if $y_1+y_2 \leq 3$, $\cro{M_P} - \cro{\M{Q}} \geq -1$. In this situation, in order to show that the max-crossing matching on $P$ has at least as many as crossings than the max-crossing matching of the corresponding $Q$, we either have to find one more crossing in $M_P$ which is not counted before, or we should define a different matching on $P$ and compare its crossings with the crossings in $\M{Q}$. If the edges emanating from red and blue points in some group of $S \subset P$ crosses each other, then we can add at least one more crossing to our counting of $\cro{M_P}$ (such crossings are not counted in the formula before). Then $\cro{M_P} \geq \cro{\M{Q}}$. Thus, the only remaining possibility is that the crossing families formed by the red and blue points in $S_{RT}$ (and $S_{LT}$) do not cross each other, i.e, $S$ has a 4--block coloring. Here, we consider the following two cases.

 		\item[Case 2.2.1:] Assume that $y_1 = 0$ and $S \subset P$ has a 4--block coloring. Then we have $x_1=x_2=y_1=0$ and $y_2 \neq 0$. If the edges emanating from $y_2$ cross any other edges that emanates from a different colored point in the same group, then $\cro{M_P} \geq \cro{\M{Q}}$. If not, then $P$ has a 4--block coloring. Then by Lemma~\ref{lem:partialsol1}, $\cro{M_P} \geq \cro{\M{Q}}$.
 		
 		\item[Case 2.2.2:]  Assume that $y_1,y_2 \neq 0$ and $S \subset P$ has a 4--block coloring.
 		If the edges emanating from any of the b-antipodal points in $P_{LT}$ crosses an edge emanating from a m-antipodal point in $S_{LT}$ of different color, then $\cro{M_P} \geq \cro{\M{Q}}$. Otherwise, we define a different matching on $P$.
 		
 		As $x_1 =x_2 =0$, $P$ has a 6--block coloring. W.l.o.g. assume that $|R_1| = 2m+1+y_1$, $|B_1|= 2m+1$, $|R_2|= y_2$, $|B_2|= y_1$, $|R_3| = 2m+1$, $|B_3| = 2m+1+y_2$. We define a matching $M'_P$ on $P$ as follows: (Note that when we say \enquote{match two collection of points}, we mean to match it as a crossing family). Match the first $m$ points in $R_1$ to the last $m$ points of $B_1$. Match next $y_1$ points of $R_1$ to $B_2$. Then match the remaining $m+1$ points of $R_1$ to the first $m+1$ points of $B_3$. Now match the next~$y_2$ points of $B_3$ to $R_2$ and the remaining~$m$ points of $B_3$ is matched to the first $m$ points of $R_3$. Next match the remaining unmatched points, that is, the first $m+1$ points in $B_1$ to the last $m+1$ points in $R_3$ (see Fig~\ref{fig:6block}). 
 		
 		\begin{figure}[htb]
 			\centering
 			\includegraphics[scale=0.5,page=1]{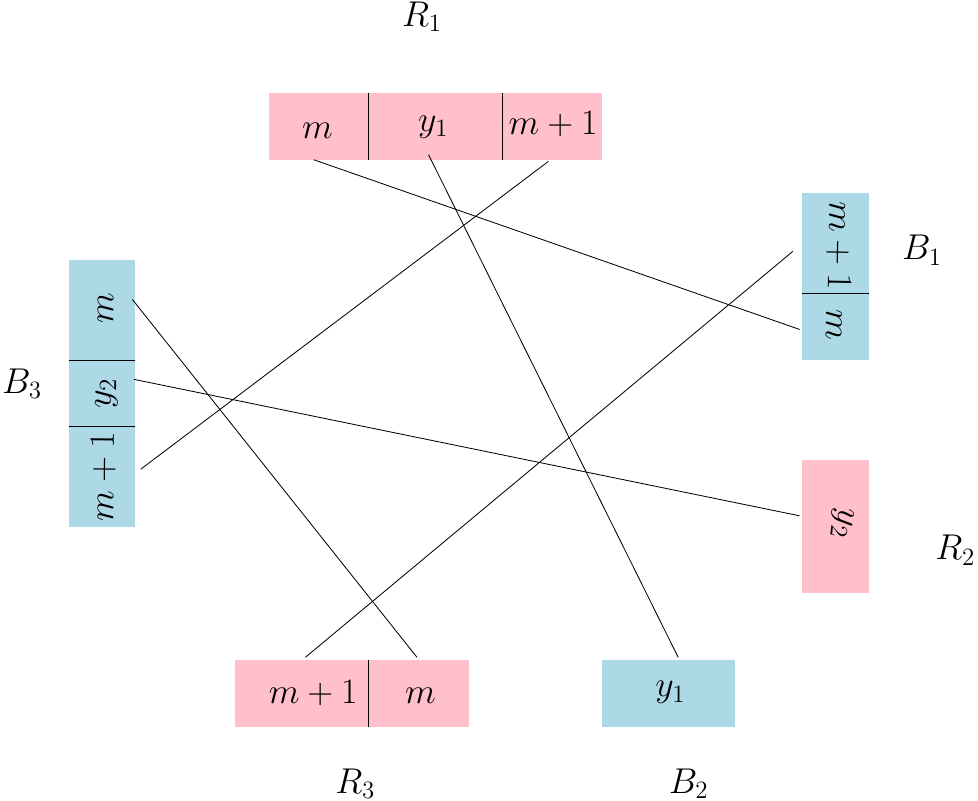}
 			\caption{ 6--blocks and their corresponding matching.}
 			\label{fig:6block}
 		\end{figure}
 	
 	The number of crossings in $M'_P = 2\binom{m}{2} +2\binom{m+1}{2} +\binom{y_1}{2} + \binom{y_2}{2} + 4m^2 + m + 2y_1(m+1)+ 2y_2(m+1)+my_1+my_2+y_1y_2$. For $y_1=y_2=1$ we get $\cro{M'_P} = 6m^2+10m+5$. The total number of red points in such a set $P$ is $n= 4m+4$, i.e.,  $c=0$. In this case, $\cro{\M{Q}} = 6m^2 +10m+4$. Thus we have $\cro{M_P} - \cro{\M{Q}} \geq 0$. For $y_1 =1$ and $y_2 = 2$, we get $\cro{M'_P} = 6m^2+13m+9$ and the total number of red points in such a set $P$ is $n= 4m+5$, i.e., $c=\frac{1}{8}$. In this case, $\cro{\M{Q}} = 6m^2 +13m+7$. Thus we have $\cro{M_P} - \cro{\M{Q}} \geq 0$. This completes the last case and hence the whole proof. 
 \end{description}
\vspace{-5ex}
\hfill\qed\end{proof}

 \newpage
\section{Proof of Proposition 1}\label{app_prop1}

\propoconvmore*

\begin{proof}
  Let $P \in \mathcal{C}_{n,n}$, where $n \geq 7$. 
	We will first partition $P$ into subsets $P_1, P_2, \ldots, P_\ell$ of 14 points each, such that each subset has 7 red and 7 blue points, 
	plus one (possibly empty) point set $P_{\ell+1}$ of at most 6 red and blue points each.
	To do this, we will iteratively split of a set $P_i$ from $P$, starting with $i=1$, in the following way:
  
	Starting with an empty set $P_i$, we traverse the points along the convex hull of $P$ in clockwise order (starting from an arbitrary point) and add them to $P_i$ one by one until $P_i$ has 14 points.
	Let $s$ be the difference between the number of red and blue points in $P_i$. 
	While $s\neq 0$, we remove the first point that was added to $P_i$ and add the next point along the convex hull of $P$. 
	In each such step, the value of $s$ either increases by 2 or decreases by 2 or remains unchanged.
    Note that since the number of red and blue points in $P$ is the same, summing the values of $s$ over all different subsets of 14 consecutive points along the convex hull of $P$ gives zero.
    Hence we eventually obtain a set $P_i$ with 7 red and 7 blue points and remove these points from $P$.
	Once the iteratively reduced set $P$ has less than 14 points, we denote this remaining (possibly empty) set as $P_{\ell+1}$.

  Recall that by computations, we know that for any $1 \le i \le \ell$ and any integer $k' \in \{0,1,2,\ldots,15\}\setminus\{1,2\}$, 
  there exists a perfect matching of $P_i$ with exactly~$k'$ crossings. 
  Further, $P_{\ell+1}$ always admits a plane perfect matching. 

  Since the convex hulls of the subsets $P_1, P_2, \ldots, P_\ell, P_{\ell+1}$ are pairwise disjoint by construction, no crossing occurs between edges from different subsets.
  Therefore, we can construct a perfect matching of $P$ with any number of crossings $k$ such that $3 \leq k \leq 15 \cdot \ell = \frac{15}{7}(n-c)$, where $c \leq 6$ is the number of red (blue, respectively) points in $P_{\ell+1}$,
  by combining appropriate matchings of the subsets $P_1, P_2, \ldots, P_\ell$. 
\hfill\qed\end{proof}

We remark that the partition of $P$ into balanced bichromatic sets with pairwise disjoint convex hulls in the above proof strongly relies on the fact that $P$ is in convex position. In fact, for a balanced bichromatic point set in general position, an according result is in general not possible. An alternative partitioning method that also works for general point sets is to recursively apply the ham sandwich theorem until all sets are sufficiently small (but not smaller than the desired value).
Note, however, that the resulting balanced bichromatic sets will have cardinalities ranging from the desired minimum value to nearly twice that value.
Using this method in the above proof would result in roughly half the desired bound for $k$.

\end{document}